\documentclass[12pt]{iopart}
\begin{document}
\jl{1}
\title{Remarks on confinement driven by axion-like particles in
Yang-Mills theories}
\author{Patricio Gaete\dag\footnote{e-mail address: patricio.gaete@usm.cl},
Euro Spallucci\ddag\footnote{e-mail address: spallucci@ts.infn.it}}
\address{\dag\ Departamento de F\'{\i}sica, Universidad T\'ecnica
F. Santa Mar\'{\i}a, Valpara\'{\i}so, Chile}
\address{\ddag\ Dipartimento di Fisica Teorica, Universit\`a di \
Trieste and INFN, Sezione di Trieste, Italy}

\begin{abstract}
Features of screening and confinement are studied for a
non-Abelian gauge theory with a mixture of pseudoscalar and scalar
coupling, in the case where a constant chromo-electric, or chromo-magnetic,
strength expectation value is present. Our discussion is carried out using the
gauge-invariant but path-dependent variables formalism.
We explicitly show that the  static potential profile is the sum of a Yukawa
and a linear potential, leading to the confinement of static probe charges.
Interestingly, similar results have been obtained in the context of
gluodynamics in curved space-time. For only pseudoscalar coupling, the results
are radically different.
\end{abstract}
\submitted
\maketitle

One of the most challenging, and still open, problem in high energy
theoretical physics is the
quantitative description of confinement in $QCD$. Recent
advances in string theory are providing
a promising new framework to face the non-perturbative features
of Yang-Mills theories (for a recent review see \cite{Nastase:2007kj}).\\
However, phenomenological models still represent a key tool
for understanding confinement physics. In this context we recall the
illustrative scenario of dual superconductivity \cite{Nambu}, where
it is conjectured that the $QCD$ vacuum behaves as a dual type $II$
superconductor. More precisely, because of the condensation of
magnetic monopoles, the chromo-electric field acting between $q
\overline q$ pairs is squeezed into strings (flux tubes), and the
nonvanishing string tension represents the proportionality constant
in the linear potential. Lattice calculations have confirmed this
picture by showing the formation of tubes of gluonic fields
connecting colored charges \cite{Capstick}.\\
With these considerations in mind, in a previous work
\cite{Gaete:2005nu}, we have studied an effective non-Abelian
gauge theory where a Cornell-like profile is obtained in the
presence of a nontrivial constant expectation value for the gauge
field strength $\langle F^{a}_{\mu\nu} \rangle$ coupled to a light
pseudoscalar boson field $\varphi$ ("axion"). In fact, this theory
experiences mass generation due to the breaking of rotational
invariance induced by the classical background configuration of the
gauge field strength, and in the case of a constant chromo-electric
field strength expectation value the static potential remains
Coulombic. Nevertheless, this picture drastically changes in the
case of a constant chromo-magnetic field strength expectation value.
In effect, the potential energy is the sum of a Coulomb and a linear
potential, leading to the confinement of static charges. It should
be noted that the magnetic character of the field strength
expectation value needed to obtain confinement is in agreement  with
the current chromo-magnetic picture of the $QCD$ vacuum
\cite{Savvidy}. Incidentally, the above static potential profile is
analogous to that encountered in Yang-Mills theory with spontaneous
symmetry breaking of scale symmetry \cite{Gaete:2007ry}. This
implies then that, although the constraint structure of the two
models is quite different, the physical content is identical. As a
result, our study has provided a new kind of ``duality'' between
effective non-Abelian theories.\\
On the other hand, in recent times the coupling of axion-like
particles with photons in the presence of an external background
electromagnetic field and its physical consequences have been object
of intensive investigations
\cite{Maiani,Sikivie,Guendelman:1991se,Stodolsky,Grifols,Jain,Ansoldi:2002id},
\cite{Masso,Heyl,Guendelman:2007cd,Guendelman:2007hb,Raffelt,Ringwald},
after recent results of the $PVLAS$ collaboration \cite{Zavattini}.
Let us also mention here that these effects can be qualitatively
understood by the existence of light pseudo-scalars bosons $\phi$
(''axions''), with a coupling to two photons. In this context, it
was suggested in \cite{Liao} that the spin-zero particle describing
the $PVLAS$ results could be one of no definite parity. The reason
is that within the low energy regime used by $PVLAS$ the spin-zero
particle could well be one of no definite parity, that is, a mixture
of pseudoscalar and scalar. Certainly, if the $PVLAS$ results are
supported by further experimental data, it would signal new physics
containing very light bosons \cite{Bibber}. Given its relevance, it
is of interest to understand better the impact of spin-zero
particle-gluon interactions on a physical observable. Seem from such
a perspective, the present work is an extension of our previous
studies started in \cite{GaeteGuen} and continued in \cite{Gaete:2005nu}. 
To do this, we will work out
the static potential for a theory which includes scalar and
pseudoscalar particles coupled to a non-Abelian gauge field using
the gauge-invariant but path-dependent variables formalism. Our
treatment is fully non-perturbative for the spin-zero field. As a
result, we obtain that the potential energy is the sum of a Yukawa
and a linear potential, leading to the confinement of static
charges, which clearly shows the key role played by the scalar
particle in transforming the Coulombic potential into the Yukawa
one. This may be contrasted with the role played by the
noncommutative space in transforming the Yukawa potential into the
Coulombic one, in the context of noncommutative axionic
electrodynamics \cite{GaeteSchmidt}. Interestingly enough, the above
static potential profile is analogous to that encountered in
gluodynamics in curved space-time \cite{Gaete:2007zn}. Therefore,
the above result reveals a new equivalence between effective
non-Abelian theories, in spite of the fact that they have different
constraint structures. Accordingly, the gauge-invariant but
path-dependent variables formalism, offers an alternative view in
which  some features of effective
non-Abelian gauge theories become more transparent. \\

We shall now discuss the interaction energy between static
point-like sources for the model under consideration. To this end we
will compute the expectation value of the energy operator $H$ in the
physical state $|\Phi\rangle$ describing the sources, which we will
denote by $ {\langle H\rangle}_\Phi$. The starting point is the
Lagrangian density:
\begin{equation}
\mathcal{L} =  - \frac{1}{4}F_{\mu \nu }^a F^{a\mu \nu }  +
\frac{1}{2}\left( {\partial _\mu  \varphi } \right)^2  -
\frac{1}{2}m^2 \varphi ^2  + \frac{{\lambda _ +  }}{4}\varphi F_{\mu
\nu }^a F^{a\mu \nu }  + \frac{{\lambda _ -  }}{4}\varphi \tilde
F_{\mu \nu }^a F^{a\mu \nu } \label{ext5}
\end{equation}
where $m$ is the mass for the spin-zero particle. Here, $A_\mu
\left( x \right) = A_\mu ^a \left( x \right)T^a$, where $T^{a}$ is a
Hermitian representation of the semi-simple and compact gauge group;
and $F_{\mu \nu }^a  = \partial _\mu  A_\nu ^a  - \partial _\nu
A_\mu ^a  + gf^{abc} A_\mu ^b A_\nu ^c$, with $f^{abc}$ the
structure constants of the group. Whereas $ \lambda_{+} $ and $
\lambda_{-} $ are couplings constants for scalar and pseudoscalar
particles, respectively. \\
Lagrangian (\ref{ext5}) provides an \textit{effective} description
of axion-like particles interacting with 
chromo-electric and chromo-magnetic fields.
Thus, from a phenomenological point of view, $m^2$ and $\lambda_\pm$
were put ``by hand'' and not related in any simple way 
to the Yang-Mills coupling constant $g$ and energy scale  $\Lambda_{QCD}$.\\
As we have indicated in \cite{GaeteGuen},
to compute the interaction energy we need to carry out the
integration over the $\varphi$-field. Once this is done, we arrive
at the following effective theory for the gauge fields:
\begin{eqnarray}
\mathcal{L}_{eff}  &=&  - \frac{1}{4}F_{\mu \nu }^a F^{a\mu \nu }  +
\frac{{\lambda _ + ^2 }}{{32}}F_{\mu \nu }^a F^{a\mu \nu }
\frac{1}{{\Delta  + m^2 }}F_{\mu \nu }^b F^{b\mu \nu } \nonumber \\
&+&\frac{{\lambda _ - ^2 }}{{32}}\tilde F^{a\mu \nu } F_{\mu \nu }^a
\frac{1}{{\Delta  + m^2 }}\tilde F^{b\mu \nu } F_{\mu \nu }^b  +
\frac{{\lambda _ +  \lambda _ -  }}{{16}}F_{\mu \nu }^a F^{a\mu \nu
} \frac{1}{{\Delta  + m^2 }}\tilde F^{b\mu \nu } F_{\mu \nu }^b
\label{ext10}
\end{eqnarray}
Next, after splitting $F_{\mu\nu}^a$ in the sum of a classical
background $\left\langle {F_{\mu \nu }^a } \right\rangle$ and a
small fluctuation $f_{\mu \nu }^a$, the corresponding Lagrangian
density becomes
\begin{equation}
\mathcal{L}_{eff} =  - \frac{1}{4}f_{\mu \nu }^a \left[ {1 -
\frac{{3\lambda _ + ^2 v^{c\lambda \rho } v_{\lambda \rho }^c
}}{{\Delta  + m^2 }}} \right]f^{a\mu \nu }  + \frac{{\lambda _ - ^2
}}{{32}}v^{a\alpha \beta } f_{\alpha \beta }^a \frac{1}{{\Delta  +
m^2 }}v^{b\gamma \delta } f_{\gamma \delta }^b \label{ext15}
\end{equation}
Here we have simplified our notation by setting $\varepsilon
 ^{\mu \nu \alpha \beta } \left\langle{F_{\mu \nu }^a } \right\rangle
\equiv v^{a\alpha \beta }$ and $\varepsilon ^{\rho \sigma \gamma
 \delta } \left\langle {F_{\rho \sigma }^b } \right\rangle
 \equiv v^{b\gamma \delta }$.\\
  We remark that the new
feature of the present model is the non-trivial presence of the
term proportional to $\lambda _ + ^2$. This point motivate us to
study the role of the scalar field on a physical observable.\\

We now turn our attention to the calculation of the interaction
energy in the $v^{a0i} \ne 0$ and $v^{aij}=0$ case (referred to as the
electric one in what follows). In such a case the Lagrangian
(\ref{ext15}) reads
\begin{equation}
\mathcal{ L}_{eff}  =  - \frac{1}{4}f_{\mu \nu }^a \left( {1 +
\frac{{6\lambda _ + ^2 \left( {v^c } \right)^2 }}{{\Delta  + m^2 }}}
\right)f^{a\mu \nu }  + v^{ai0} f_{i0}^a
\frac{{{\raise0.7ex\hbox{${\lambda _ - ^2 }$} \!\mathord{\left/
 {\vphantom {{\lambda _ - ^2 } 8}}\right.\kern-\nulldelimiterspace}
\!\lower0.7ex\hbox{$8$}}}}{{\Delta  + m^2 }}v^{bk0} f_{k0}^0
\label{ext20}
\end{equation}

The canonical Hamiltonian can be worked as usual and is given by
\begin{eqnarray}
H_C  &=& \int {d^3 x} \left[ {\Pi ^{ai} \left( {\partial _i A_0^a +
gf^{abc} A_0^c A_i^b } \right) + \frac{1}{2}B^{ai} \left( {1 +
\frac{{6\lambda _ + ^2 \left( {v^c } \right)^2 }}
{{\Delta  + m^2 }}} \right)B^{ai} } \right]  \nonumber \\
&+& \int {d^3 x} \left[ {\frac{1}{2}\Pi ^{ai} \frac{{\Delta  + m^2
}} {{\Delta  + M^2 }}\Pi ^{ai}  - \frac{{\lambda _ - ^2 }}{8} \left(
{v^{ai} \Pi ^{ai} } \right)\frac{1}{{\Delta  + \mathcal{ M}^2 }} \left(
{v^{bi} \Pi ^{bi} } \right)} \right]  \label{ext35}
\end{eqnarray}
where $B^{ai}$ is the chromo-magnetic field,  $M^2  \equiv m^2  +
6\lambda _ + ^2 \left( {v^a } \right)^2$, and $\mathcal{ M}^2
\equiv M^2  + {\raise0.7ex\hbox{${\lambda _ - ^2 }$}
\!\mathord{\left/
 {\vphantom {{\lambda _ - ^2 } 4}}\right.\kern-\nulldelimiterspace}
\!\lower0.7ex\hbox{$4$}}\left( {v^a } \right)^2  = m^2  + \left[
{6\lambda _ + ^2  + {\raise0.7ex\hbox{${\lambda _ - ^2 }$}
\!\mathord{\left/
 {\vphantom {{\lambda _ - ^2 } 4}}\right.\kern-\nulldelimiterspace}
\!\lower0.7ex\hbox{$4$}}} \right]\left( {v^a } \right)^2$.\\
By proceeding in the same way as in \cite{Gaete:2005nu}, we obtain
 the  static potential for two opposite charges
located at ${\bf 0}$ and ${\bf y}$:
\begin{equation}
V =  - \frac{{g^2 }}{{4\pi }}C_F \frac{{e^{ - ML} }}{L} + g^2 \left(
{\xi  + g^2 \xi ^ \prime  } \right)L \label{ext120}
\end{equation}
where
\begin{equation}
\xi  \equiv \frac{1}{{2\pi }}\left[ {\frac{{m^2 }}{4}C_F \ln
\left( {1 + \frac{{\overline \Lambda  ^2 }}{{M^2 }}} \right) -
\frac{{\lambda _ - ^2 }}{{16}}tr\left( {v^{ai} T^a v^{bi} T^b }
\right)\ln \left( {1 + \frac{{\tilde \Lambda ^2 }}{{M^2 }}} \right)}
\right] \label{ext125}
\end{equation}
and
\begin{eqnarray}
\xi ^ \prime   &\equiv& tr\left( {f^{abc} f^{adc} T^b T^d }
\right)\left[ {\frac{1}{{8\pi }}\left( {\Lambda ^2  - M^2 \ln \left(
{1 + \frac{{\Lambda ^2 }}{{M^2 }}} \right)} \right) + \frac{{m^2
}}{4}\ln \left( {1 + \frac{{\overline \Lambda  ^2 }}{{M^2 }}}
\right)} \right] \nonumber\\
&-& \frac{{\lambda _ - ^2 }}{{16}}tr\left( {v^{pi} f^{pbc} T^b
v^{qi} f^{qdc} T^d } \right)\ln \left( {1 + \frac{{\tilde \Lambda ^2
}}{{\mathcal{ M}^2 }}} \right) \label{ext130}
\end{eqnarray}

where, $\overline \Lambda$ and $\tilde \Lambda$ are cutoffs.\\
Expression (\ref{ext120}) immediately shows both expected and unexpected
features of the model. The linear confining piece was expected from
our previous study \cite{Gaete:2005nu}.
The novel feature is the Yukawa piece. In fact, the above result clearly
reveals the key role played by the scalar particle
(${\lambda _ +  }$ term) in transforming the Coulombic potential into the
Yukawa one. As already expressed, similar form of interaction potential
has been reported before in the context of gluodynamics in curved
space-time \cite{Gaete:2007zn}. Also, a common feature of these models is
that the rotational symmetry is restored in the resulting interaction energy.\\

Now we focus on the case $v^{a0i} = 0$ and $v^{aij} \ne 0$ which we refer to
as the magnetic one in what follows. Thus, we obtain from (\ref{ext15})

\begin{equation}
\mathcal{ L}_{eff}  =  - \frac{1}{4}f_{\mu \nu }^a \left( {1 +
\frac{{6\lambda _ + ^2 \left( {v^c } \right)^2 }}{{\Delta  + m^2 }}}
\right)f^{a\mu \nu }  + v^{aij} f_{ij}^a
\frac{{{\raise0.7ex\hbox{${\lambda _ - ^2 }$} \!\mathord{\left/
 {\vphantom {{\lambda _ - ^2 } 8}}\right.\kern-\nulldelimiterspace}
\!\lower0.7ex\hbox{$32$}}}}{{\Delta  + m^2 }}v^{bkl} f_{kl}^0
,\label{ext140}
\end{equation}
where $\mu,\nu=0,1,2,3$ and $i,j,k,l=1,2,3$. Here again, the quantization is
carried out using  Dirac's procedure. The canonically conjugate momenta,
as obtained from (\ref{ext140}), are
\begin{equation}
\Pi ^{ao}  = 0, \label{ext145a}
\end{equation}
\begin{equation}
\Pi _i^a  = D_{ij}^{ab} f_{j0}^b,\label{ext145b}
\end{equation}
\begin{equation}
D_{ij}^{ab}  \equiv \delta ^{ab} \left( {1 + \frac{{6\lambda _ + ^2
\left( {v^c } \right)^2 }}{{\Delta  + m^2 }}} \right)\delta _{ij}
\label{ext145c}
\end{equation}
The explicit form of the chromo-electric field turns out to be

\begin{equation}
E_i^a  =  \frac{{\Delta  + m^2 }}{{\Delta  + M^2
}} \Pi _i^a, \label{ext50}
\end{equation}
where $M^2  \equiv m^2  + 6\lambda _ + ^2 \left( {v^a } \right)^2$.
This leads us to the canonical Hamiltonian
\begin{eqnarray}
H_C  &=& \int {d^3 x} \left[ {\Pi ^{ai} \left( {\partial _i A_0^a +
gf^{abc} A_0^c A_i^b } \right) + \frac{1}{2}B^{ai} \left( {1 +
\frac{{6\lambda _ + ^2 \left( {v^c } \right)^2 }}
{{\Delta  + m^2 }}} \right)B^{ai} } \right]  \nonumber \\
&+& \int {d^3 x} \left[ {\frac{1}{2}\Pi ^{ai} \frac{{\Delta  + m^2
}} {{\Delta  + M^2 }}\Pi ^{ai}} \right]\ ,  \label{ext155}
\end{eqnarray}
where $B^{ai}$ is the chromo-magnetic field. \\
We skip all the technical details and refer to \cite{Gaete:2005nu} for
them.   The static potential turns out to be
\begin{equation}
V =  - \frac{{g^2 }}{{4\pi }}C_F \frac{{e^{ - ML} }}{L} +
g^2 \left( {\xi  + g^2 \xi ^ \prime  } \right)L, \label{ext180}
\end{equation}
where
\begin{equation}
\xi  = \frac{{m^2 }}{{8\pi }}C_F
\ln \left( {1 + \frac{{\overline \Lambda  ^2 }}{{M^2 }}} \right),
\label{ext185}
\end{equation}
and
\begin{equation}
\xi ^ \prime   =
\frac{1}{4}\left[ {\frac{{C_F C_A \sigma }}{{2\pi ^2 }} +
m^2 tr\left( {f^{abc} T^b f^{adc} T^d } \right)
\ln \left( {1 + \frac{{\overline \Lambda  ^2 }}{{M^2 }}} \right)} \right]
\label{ext190}
\end{equation}

Here, in contrast to our previous analysis \cite{Gaete:2005nu},
unexpected features are found. Interestingly, it is observed that
the introduction of the ${\lambda _ +  }$ term induces a Yukawa
piece plus a linear confining piece. Notice that in the ${\lambda _
+ =0 }$  case, the static potential remains Coulombic. Again, the
rotational symmetry is restored in the resulting interaction energy
despite the chromo-electric external field breaks this
symmetry. \\
The presence of mass terms, coming from a non vanishing
background value for $F^{a}_{\mu\nu}$, suggests a possible analogy
with the model introduced in \cite{Vercauteren:2007gx}, where the
gluon mass is produced through a non-vanishing vacuum expectation
value of the composite operator $A^2 _{\mu}$. However, our approach
is substantially different: the distinctive feature of our method is
to define the interaction potential between test charges in a
manifestly gauge invariant way. On the other hand, the non-vanishing
vacuum expectation value $\langle \, A^2_{\mu}\,\rangle$ in
\cite{Vercauteren:2007gx} is gauge dependent. The authors claim that
under some circumstances this quantity can be given a gauge
invariant meaning, and we trust them, but a direct comparison with
our approach is very hard, at this level. A second difference it can
be worth to remark is the different kind of "energies" which are
considered. The main purpose of \cite{Vercauteren:2007gx} is to
resolve the instability of the Saviddy vacuum. Accordingly, the
authors compute one-loop effective potential, i.e. vacuum energy
density, in the presence of a background chromomagnetic field and a
condensate for $A^2_{\mu}$. On our side, we determine the
interaction energy between static charges, which is the static
potential energy. Comparison with \cite{Vercauteren:2007gx} would
require the calculation of the one-loop vacuum energy density for
our model, but this is a problem which is interesting by itself and
cannot be fully investigated in this short note.\\
Let us put our work in its proper perspective. This paper is a sequel
to Ref. \cite{GaeteGuen, Gaete:2005nu}, where we have exploited a crucial point for
understanding the physical content of gauge theories, that is, the
identification of field degrees of freedom with observable quantities.
 Our analysis reveals both expected and unexpected features of the model
 studied. It was shown that the static potential profile is the sum of a
 Yukawa and a linear potential, leading to the confinement of static probe
 charges. This result is obtained for both external chromo-magnetic and
 chromo-electric strength expectation value. This may be contrasted with our
 previous study Ref. \cite{Gaete:2005nu}, where only a pseudoscalar
 coupling was considered. Also, the above analysis it has showed the key role
 played by the scalar particle in transforming the Coulombic potential into
 the Yukawa one.
Interestingly, similar results have been obtained in the context of
gluodynamics on curved space-time  \cite{Gaete:2007zn}. An important
consequence of this is that, although the constraint structure of
the two models is quite different, the physical content is
identical. This means that our study has provided a new kind of
``duality'' between effective
non-Abelian theories which is summarized in the picture below.\\
We conclude noting that our results agrees with the dilaton coupled to gauge
fields mechanism \cite{Levin}. However, although both approaches lead to
confinement, the above analysis reveals that the mechanism of obtaining a
linear potential is quite different. As already mentioned, in this work we
have exploited the similarity between the tree level mechanism that leads to
confinement here and the nonperturbative mechanism (caused by quantum effects)
which gives confinement in $QCD$ on curved space-time.\\

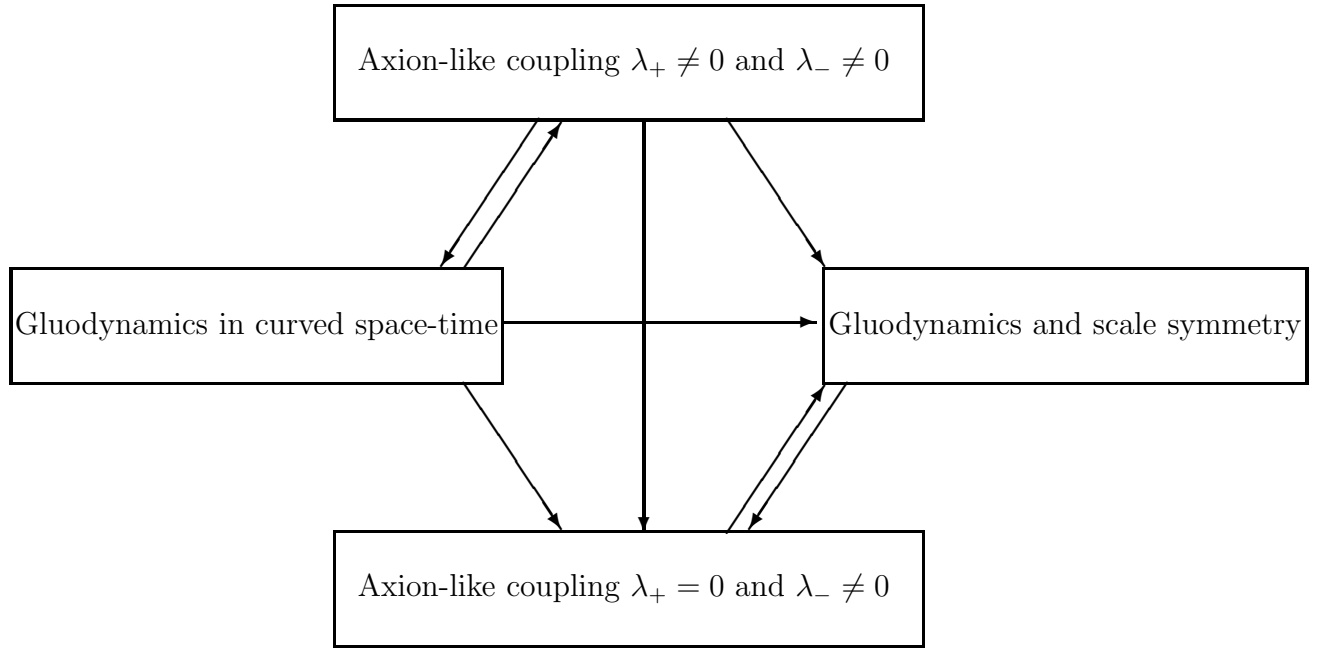
\setlength{\unitlength}{1.0cm}\thicklines
\begin{picture}(9,11)
\put(4.3,1.5){\framebox(7.8,1.5){Axion-like coupling ${\lambda _ +  =0}$ and ${\lambda _ -\neq0 }$ }}
\put(9.5,3.0){\vector(2,3){1.3}}
\put(11.1,5.0){\vector(-2,-3){1.3}}
\put(4.3,8.5){\framebox(7.8,1.5){Axion-like coupling ${\lambda _ +  \neq0}$ and ${\lambda _ -\neq0 }$ }}
\put(6.0,6.5){\vector(2,3){1.3}}
\put(8.4,8.5){\vector(0,-3){5.5}}
\put(0.0,5.0){\framebox(6.5,1.5){Gluodynamics in curved space-time}}
\put(6.0,5.0){\vector(2,-3){1.3}}
\put(6.5,5.8){\vector(4,0){4.19}}
\put(10.8,5.0){\framebox(6.4,1.5){Gluodynamics and scale symmetry}}
\put(9.5,8.5){\vector(2,-3){1.3}}
\put(7.0,8.5){\vector(-2,-3){1.3}}
\end{picture}

Acknowledgments\\

P. G. was partially supported by Fondecyt (Chile) grant 1050546. One
of us (PG) wants to thank the Physics Department of the Universit\`a
di Trieste for hospitality.\\

\end{document}